\documentclass[12pt,preprint]{aastex}

\usepackage{epsfig}
\usepackage{amsmath}
\usepackage{amssymb}
\usepackage{natbib}
\usepackage{graphicx}

\shorttitle{The size evolution of ellipticals}
\shortauthors{Naab et al.}

\begin{document}

\title{Minor mergers and the size evolution of elliptical galaxies}

\author{Thorsten Naab$^1$, Peter H. Johansson$^1$, Jeremiah P. Ostriker$^2$}
\affil{$^1$ Universit\"ats-Sternwarte M\"unchen, Scheinerstr.\ 1, D-81679 M\"unchen,  
Germany; $^2$ Department of Astrophysics, Peyton Hall, Princeton, USA\texttt{naab@usm.lmu.de} \\}

\begin{abstract}
Using a high resolution hydrodynamical cosmological simulation of the formation of 
a massive spheroidal galaxy we show that elliptical galaxies can be very compact 
and massive at high redshift in agreement with recent observations. Accretion of 
stripped in-falling stellar material increases the size of the system with time and the 
central concentration is reduced by dynamical friction of the surviving stellar cores. 
In a specific case of a spheroidal galaxy with a final stellar mass of $1.5 \times 10^{11} M_{\odot}$  
we find that the effective radius $r_e$ increases from $0.7 \pm 0.2 \ \rm kpc$
at z = 3 to $r_e = 2.4 \pm 0.4 \ \rm kpc$ 
at z = 0 with a concomitant decrease in the effective density of an order of magnitude and a decrease of the 
central velocity dispersion by approximately 20\% over this time interval.
A simple argument based on the virial theorem shows that during the accretion of weakly bound 
material (minor mergers) the radius can increase as the square of the mass in contrast 
to the usual linear rate of increase for major mergers. By undergoing minor mergers compact high redshift 
spheroids can evolve into present-day systems with sizes and concentrations similar to 
observed local ellipticals. This indicates that minor mergers may be the  main driver for 
the late evolution of sizes and densities of early-type galaxies. 
\end{abstract}


\section{Introduction}
There is recent observational evidence that a significant fraction of 
massive evolved spheroidal stellar systems is already in place at redshift $z \ge 2$. 
However, only a small percentage of these galaxies is fully assembled \citep{2008ApJ...677L...5V}.   
The galaxies are smaller by a factor of three to five compared to present-day 
ellipticals at similar masses \citep{2005ApJ...626..680D,2007MNRAS.374..614L,2007ApJ...671..285T,
2007MNRAS.382..109T,2008ApJ...677L...5V,2008A&A...482...21C,2008arXiv0810.2795S}. 
Their effective stellar mass densities are at least  
one order of magnitude higher \citep{2008ApJ...677L...5V,2009ApJ...695..101D} with
significantly higher surface brightnesses  compared to their low redshift analogs 
\citep{2008A&A...482...21C}. 

These observations are difficult to reconcile with some current 
idealized formation scenarios for elliptical galaxies. A simple conclusion from the data 
is that most early-type galaxies can neither have fully formed in a simple monolithic 
collapse nor a binary merger of gas-rich disks at high redshift, unless their 
increase in size can be explained by secular processes such as adiabatic expansion 
driven by stellar mass loss and/or strong feedback \citep{2009ApJ...695..101D, 2008ApJ...689L.101F}. Additionally,  
simple passive evolution of the stellar population is in contradiction 
with observations of local ellipticals \citep{2008ApJ...677L...5V}. \\

Dry (i.e. gas-poor, collisionless) mergers and stellar accretion events  
are the prime candidates for the strong mass and size evolution of stellar spheroids 
at $z <2$  \citep{2006ApJ...636L..81N,2006ApJ...648L..21K,
2006ApJ...640..241B,2006ApJ...652..270B,2009arXiv0902.0373R,2009ApJ...691.1424H,2009arXiv0903.4857V} 
as the additional presence of a dissipative component in a major merger event 
would limit the size increase (see e.g. \citealp{2007ApJ...658...65C}). The observed ellipticals are already very massive 
at high redshift, thus we expect from the shape of the mass function that minor mergers should be much
more common than major mergers until z=0. 

Massive early-type galaxies may undergo not more than one 
major merger (with typically low cool gas content, see also \citealp{2008ApJ...676L.105W}) since 
$z=0.7$ (\citealp{2006ApJ...640..241B}, see also \citealp{2008MNRAS.388.1537M}) with a 
significant contribution from minor mergers for the mass buildup \citep{2009arXiv0902.1188B}. 
The low number of observed major early-type mergers is also supported by theoretical evidence that massive 
($\approx 10^{11-12} M_{\odot}$) halos at $z=2$ typically  experience only one major merger or less 
until $z=0$ and minor mergers are much more common 
\citep{2008ApJ...688..789G,2008arXiv0809.1734K}. On average, this is not enough 
to account for the required mass and size growth (see also \citealp{2009ApJ...695..101D}) as major 
dry mergers at most increase the size of a simple 
one component system by a factor of two and allowing for dark matter halos
reduces the size growth further \citep{2003MNRAS.342..501N,2006MNRAS.369..625N,2008ApJ...679..156H}.  

In this Letter we use, as a proof of principle, a very high resolution cosmological simulation 
of the formation of a spheroid with no major mergers below $z=3$ in combination with simple scaling relations 
to show that the observed rapid size growth and density evolution of spheroidal galaxies can be  
explained by minor mergers and small accretion events. The problem is computationally 
very expensive. At high redshift the observed ellipticals have half-mass sizes
of $< 1 \ \rm{kpc}$ with accreting 
subsystems of even smaller size. As we know from isolated merger simulations 
(see e.g. \citealp{2003ApJ...597..893N}), to resolve such a system reasonably well we 
require a force softening of 10\% of the effective radius, 
which in our case is of the order of $100 \ \rm pc$ and we require particle numbers of 
$\approx 10^7$ to simulate the galaxy in a full cosmological context over a
Hubble time. Finally, to accurately follow the kinematics high force and integration accuracy are required.


\section{Size evolution and minor mergers}

Using the virial theorem we make a simple estimate of how an initial one-component 
stellar systems evolves when mass in stellar systems is added. We assume that a 
compact initial stellar system has formed dissipatively from stars.  This system 
has a total energy $E_i$, a mass $M_i$, a gravitational radius $r_{g,i}$, and the mean 
square speed of the stars is $\langle v_i^2 \rangle $. According to the virial theorem 
\citep{2008gady.book.....B} the total energy of the system is 

\begin{eqnarray} 
E_i & = &  K_i+W_i = -K_i = \frac{1}{2} W_i \nonumber \\
    & = & -\frac{1}{2} M_i \langle v_i^2 \rangle = -\frac{1}{2} \frac{GM_i^2}{r_{g,i}}.
\end{eqnarray}

We then assume that systems are accreted with energies totaling $E_a$, masses totaling $M_a$,  
gravitational radii $r_{a,i}$ and mean square speeds averaging $\langle  v_a^2\rangle $. We define the 
fractional mass increase from all the accreted material $\eta = M_a/M_i$ and the total kinetic energy of 
the material as $K_a=(1/2) M_a \langle v_a^2\rangle$, further defining 
$\epsilon = \langle v_a^2 \rangle/\langle v_i^2\rangle $. Assuming energy conservation (orbital parameters
from cosmological simulations \citep{2006A&A...445..403K} indicate that most halos merge on parabolic orbits), 
the total energy of the final system is 

\begin{eqnarray}
E_f & = & E_i+E_a=-\frac{1}{2} M_i\langle v_i^2\rangle  - \frac{1}{2}M_a\langle v_a^2\rangle \nonumber \\
    & = & -\frac{1}{2} M_i\langle v_i^2\rangle  - \frac{1}{2}\eta M_i \epsilon \langle v_i^2\rangle  \nonumber \\
& = & - \frac{1}{2} M_i \langle v_i^2\rangle (1+\epsilon\eta) \nonumber \\
   & = & -\frac{1}{2} M_f \langle v_f^2\rangle .
\end{eqnarray}

The mass of the final system is $M_f = M_i + M_a= (1+\eta)M_i$. Therefore the ratio of the 
final to initial mean square speeds is 

\begin{equation}
\frac{\langle v_f^2\rangle }{\langle v_i^2\rangle } = \frac{(1+\eta\epsilon)}{1+\eta}. \label{disp}
\end{equation}

Similarly, the ratio of the final to initial gravitational radius is

\begin{equation} 
\frac{r_{g,f}}{r_{g,i}} = \frac{(1+\eta)^2}{(1+\eta\epsilon)} \label{rg}
\end{equation}

and for the ratio of the densities we get 

\begin{equation}
\frac{\rho_f}{\rho_i} = \frac{(1+\eta \epsilon)^3}{(1+\eta)^5}. \label{dens}
\end{equation}

If during one or more mergers the initial stellar system increases its 
mass by a factor of two then $\eta = 1$. This mass increase can be caused by 
one equal-mass merger in which case the mean square velocities of the two systems 
are identical and remain unchanged in the final system (Eqn. \ref{disp}). The radius 
increases by a factor of two (Eqn. \ref{rg}) and the density drops by a factor of 
four (Eqn. \ref{dens})(see also \citealp{2009ApJ...691.1424H}). If, however, the total mass increase by a factor of two is 
caused by accretion of very small systems with $\langle v_a^2\rangle  << \langle v_i^2\rangle $ or $\epsilon << 1$, 
then the mean square velocities are reduced by a factor two, the radius is four times larger and the 
density is reduced by a factor of 32 with respect to the initial system (see also \citet{2009ApJ...697.1290B} 
for a similar derivation of the scaling relations). We know from the 
shape of the Schechter function for the distribution of stellar masses that a massive system 
($m > M_*$) accretes most of its mass from lower mass systems and thus
the simple calculation above makes it very plausible that even though major mergers do occur
minor mergers are the main driver for the evolution 
in size and density of massive galaxies.   

\begin{figure}
\centering 
\includegraphics[width=8cm]{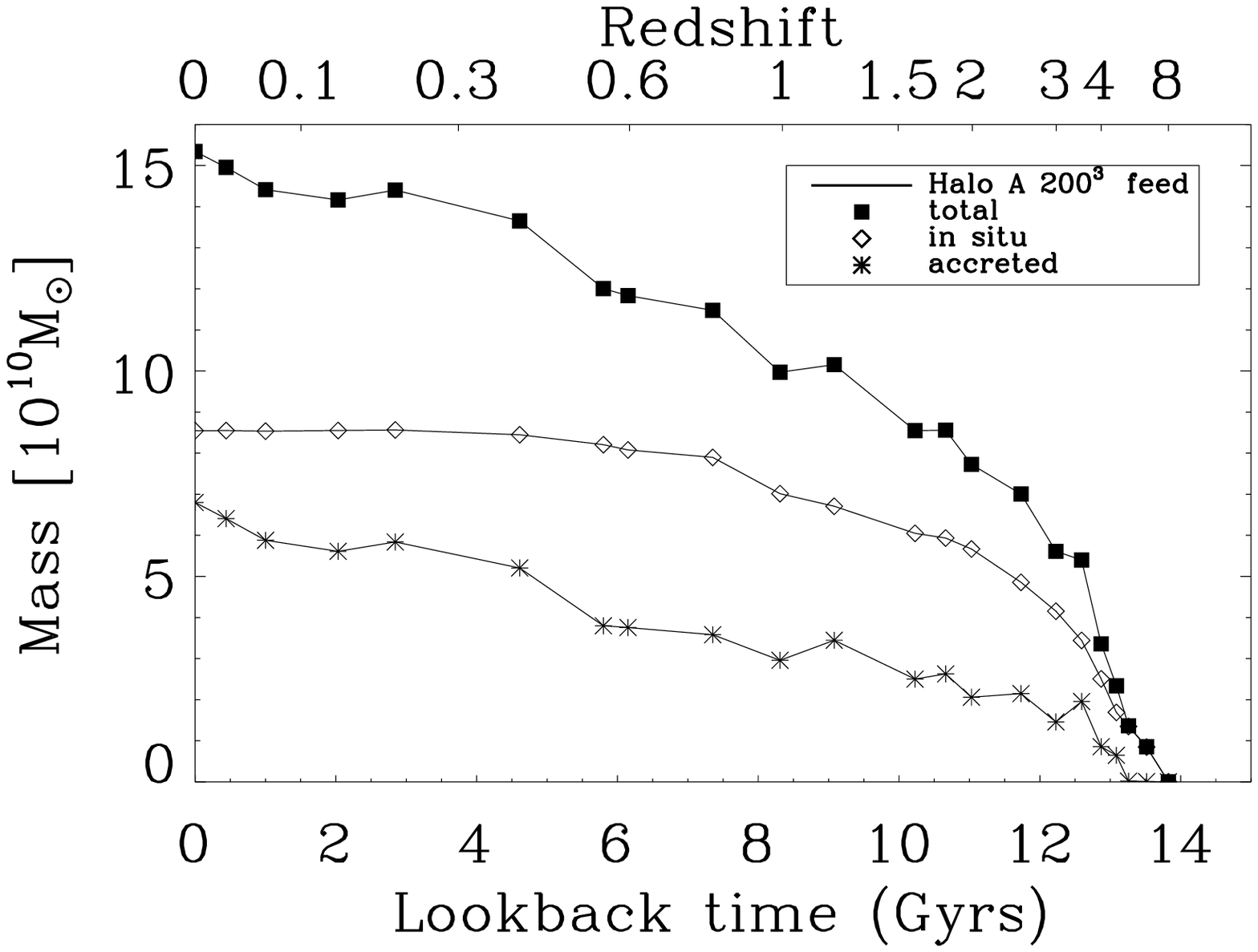}
\caption{Mass assembly history of the stellar system (squares) separated into stars made in-situ (open diamonds) in the galaxy 
and stars formed outside the galaxy that have been accreted (stars) later on. At high redshift ($z>2$) the system assembles 
by the formation of in-situ stars, at low redshift ($z<1$) accretion is more dominant.}
\label{new}
\end{figure}

\section{High resolution simulations of an individual galaxy halo} 
\label{SIMULATIONS}

We have performed a cosmological N-body/SPH high-resolution re-simulation of 
an individual galaxy halo. The process of setting up the initial conditions is 
described in detail in \citet{2007ApJ...658..710N}
and is briefly reviewed. We have re-run galaxy A at $200^3$ particles resolution 
using a WMAP-1 \citep{2003ApJS..148..175S} cosmology with a 
Hubble parameter of $h=0.65$ ($\equiv H_{0}$=100$h$ kms$^{-1}$Mpc$^{-1}$) 
with $\sigma_8$=0.86, $f_{b}= \Omega_b/\Omega_m$=0.2, $\Omega_0$=0.3, and 
$\Lambda_0$=0.7. To re-simulate the target halo at high resolution we 
increased the particle number to $1.6 \times 10^7$ gas and dark matter particles within
a cubic volume at redshift $z=24$ containing all particles that end up
within the virialized region (we assumed a fixed radius of $0.5 \ \rm Mpc$) 
of the halo at $z=0$. The tidal forces from particles outside the high resolution cube 
were approximated by increasingly massive dark matter particles in 5 nested 
layers of decreasing resolution. The galaxy was not contaminated by massive boundary particles 
within the virial radius. 

\begin{figure}
\centering 
\includegraphics[width=8cm]{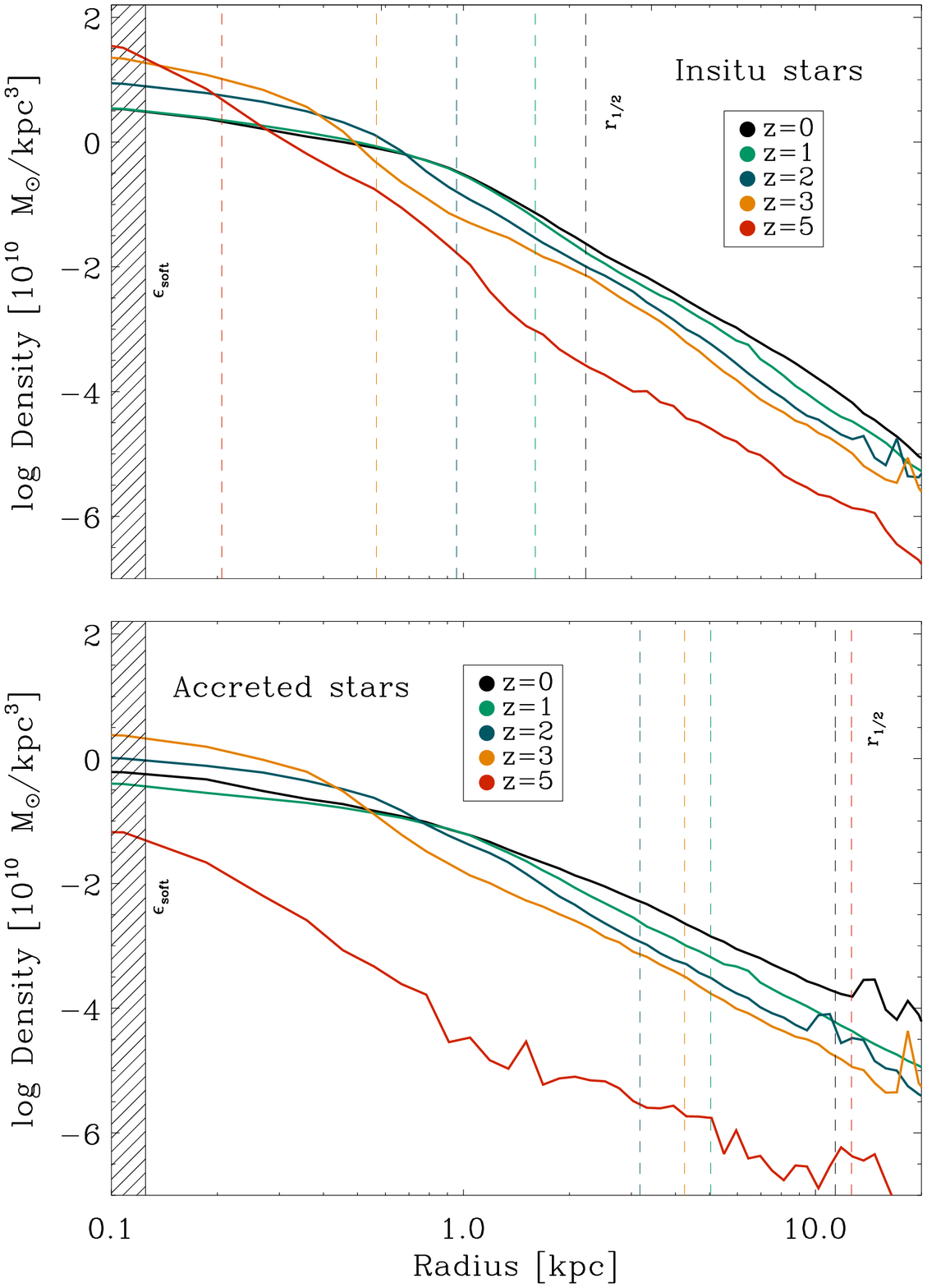}
\caption{Density profile of the stars formed in-situ in the galaxy (upper plot) and of stars 
formed outside the galaxy and then accreted later-on (bottom plot) at redshifts z=5,3,2,1,0 
(red,orange,blue,green,black). The spherical half mass radii $r_{1/2}$ are indicated by the dashed vertical 
lines. The shaded area indicates the gravitational softening length.}
\label{rho_ins_acc_feed_200_comb}
\end{figure}

The simulation was performed with GADGET-2 \citep{2005MNRAS.364.1105S} on Woodhen
at the Princeton PICSciE HPC center using a total of 177,000 CPU hours on 64 CPUs. 
We used a fixed comoving softening until $z=9$, 
and thereafter the softening, e.g. for the stars, remained fixed at physical $125 \rm pc$.
The mass of an individual stellar particle is $1.3 \times 10^5 M_{\odot}$ and we spawn two stars per
SPH particle. 

Star formation and feedback from supernovae was included using 
the sub-grid multiphase model of \citep{2003MNRAS.339..289S}. We require an 
over-density contrast of $\Delta > 55.7$ for the onset of star formation 
to avoid spurious star formation at high redshift. The threshold number 
density for star formation is $n_{\rm thresh} = 0.205 \ \rm cm^{-3}$ and the star formation 
time-scale is $t_* = 1.5 h^{-1} \rm Gyrs$. We also included an uniform UV background 
radiation field peaking at at $z\simeq 2-3$ (see \citealp{2007ApJ...658..710N}). 

At present the galaxy has a total virial mass of $1.9 \times 10^{12} M_{\odot}$ and a total 
stellar mass of $1.5 \times 10^{11} M_{\odot}$. The ratio of central stellar mass to halo mass 
is about a factor of two larger than predicted from gravitational lensing studies 
\citep{2006MNRAS.368..715M}, however, it is comparable to some recent predictions 
derived for the Milky Way halo ( e.g. \citealp{2008ApJ...684.1143X}, 
see however \citealp{2008MNRAS.384.1459L}).  The central stellar component 
resembles an elliptical galaxy with properties very similar to the results presented
in \citet{2007ApJ...658..710N} and in this letter we only focus on particular aspects of 
the assembly and size evolution of the stellar component. 

\section{Results}
\begin{figure}
\centering 
\includegraphics[width=8cm]{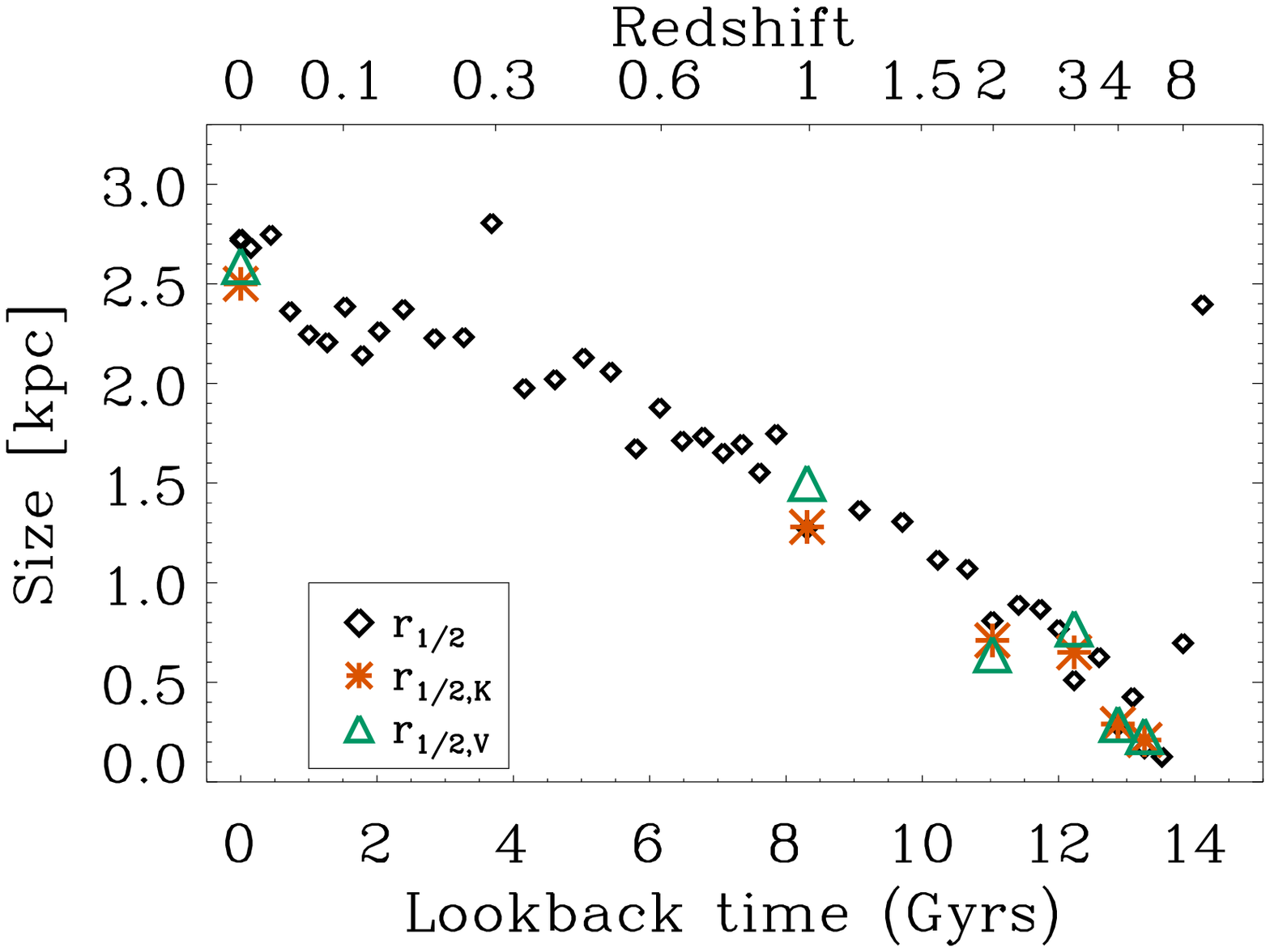}
\caption{Time evolution of the edge-on circular projected stellar half-mass radius within 
fixed physical $30 \ \rm kpc$ (black diamonds). From $z=3-2$ to $z=0$ the size increases by 
a factor of $\approx 3-4 $.We also show the evolution of the rest-frame K-band 
(red cross) and V-band (green triangle) half-light radius. Outliers
indicate minor merger events, e.g. the most massive (8:1) merger since z=3 at z=0.3.}
\label{re}
\end{figure}

During the assembly of the central galaxy we have separated the stars 
within a fiducial radius of $30 \ \rm kpc$ in fixed physical coordinates into stars that have 
formed in-situ from gas within the galaxy and stars that have formed outside this radius 
and were accreted later-on. The mass assembly history of the in-situ and accreted 
components of the stellar system are shown in Fig. \ref{new}. The early mass evolution at $z >2$ is driven by 
the assembly of in-situ stars with a decreasing contribution towards $z \approx 0.7$.
Below this redshift only few stars are formed within $30 \ \rm kpc$. The final 20\% 
of stars are added thereafter by accretion of systems formed outside the main stellar 
system at radii larger than $30 \ \rm kpc$.    

The upper panel of Fig. \ref{rho_ins_acc_feed_200_comb} shows the density profiles of the in-situ stars 
at redshifts z=5,3,2,1,0. Between z=5 and z=3 the central galaxy is still building 
up from gas flows feeding the central region of the galaxy directly, forming 
a concentrated stellar system. The in-situ central stellar densities decrease by more than an order 
of magnitude towards lower redshifts. 
The spherical half-mass radii of the in-situ stellar component, show that the in-situ system is very 
compact (see also \citealp{2009ApJ...692L...1J}) at $z=3$ $(0.5 - 0.6 \ \rm kpc)$ and its size increases by about a 
factor of four ($\approx 2 \ \rm kpc$) until z=0. In the bottom panel of Fig. \ref{rho_ins_acc_feed_200_comb} we 
show the density profiles for the stars that have formed outside $30 \ \rm kpc$ and have been accreted later-on. 
This component is more extended at all redshifts and has a shallower density profile. 
Its central density stays almost constant at $\approx 10^{10} M_{\odot} \rm kpc^{-3}$ while the density at 
larger radii subsequently increases towards $z=0$. The half-mass radius of this component 
is significantly larger than for the in-situ stars ($> 3 \ \rm kpc$). The central part of the galaxy is always 
dominated by in-situ stars whereas at redshifts below $z \approx 2-3$ and at radii larger than 
$\approx 2 -3 \ \rm kpc$ the system is dominated by accreted stars. 

\begin{figure}
\centering 
\includegraphics[width=8cm]{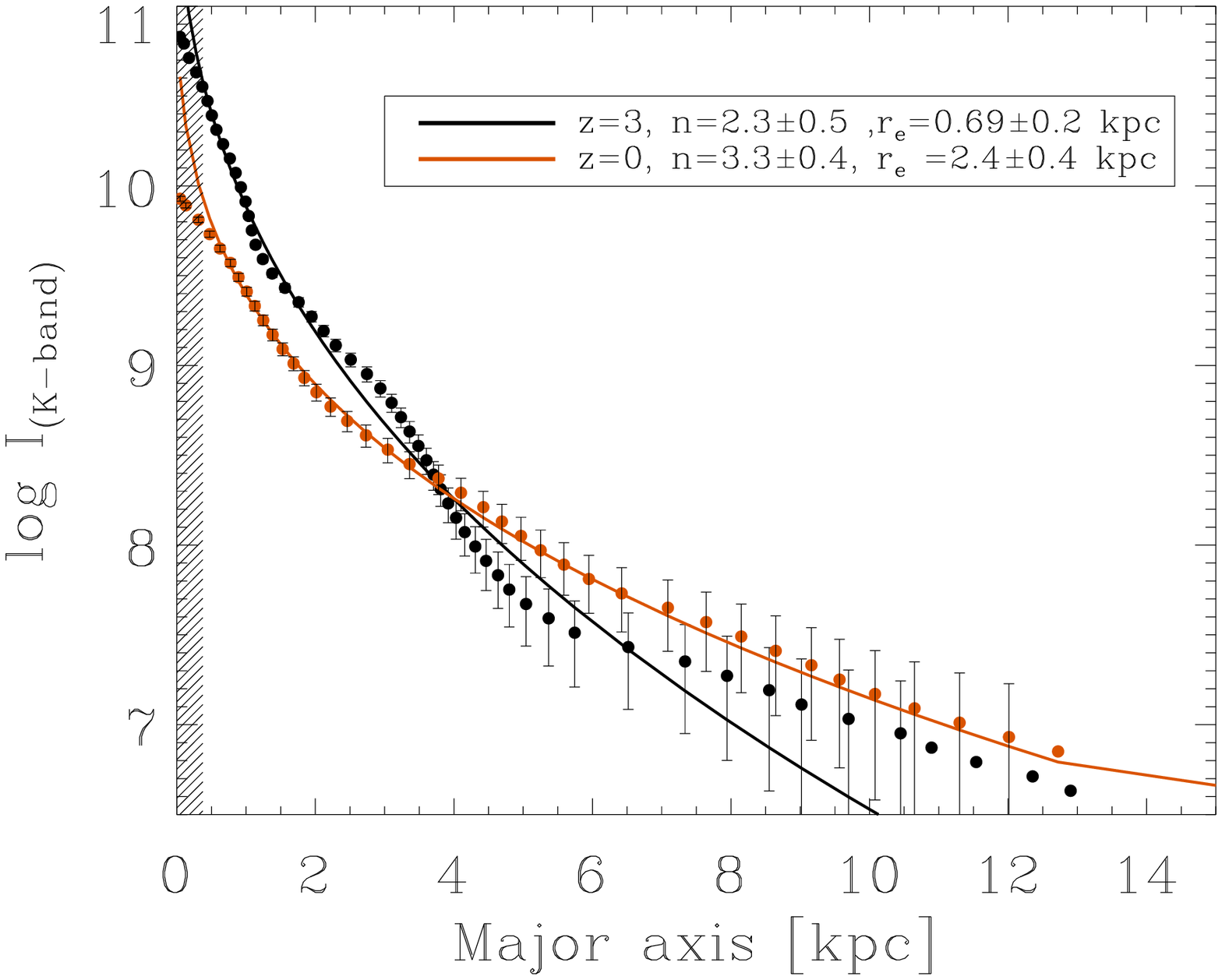}
\caption{K-band surface brightness profiles assuming solar metalicity at $z=3$ (black dots) 
and $z=0$ (red dots). The black and red lines indicate the best fitting Sersic profile. At high 
redshift the galaxy has a higher central surface brightness and is more compact.}
\label{kmag_all}
\end{figure}

In Fig. \ref{re} we show the time evolution of the edge-on projected half mass radius of stars 
in the central galaxy within the central physical $30 \ \rm kpc$ as a function of time. 
At $z=3$ the stellar system resembles a compact disk-like or bar-like object with a peak 
ellipticity of $\epsilon =0.65$ and a size of $\approx 0.3 - 0.7 \ \rm kpc$ at z=3. Thereafter its 
size increases by a factor of $\approx 3-4$ to its present value of $2.7 \ \rm kpc$. We also plot 
the projected half-light radii in the rest frame K- and  V-band using the stellar population models of 
\citet{2003MNRAS.344.1000B} assuming solar metalicity. In general the half-mass radii trace 
the half-light radii even at larger redshifts reasonably well.

The K-band rest-frame surface brightness profiles for edge-on projections 
at z=0 and z=3 are shown in Fig. \ref{kmag_all} in combination with the best fitting Sersic profiles.
using the fitting procedure of \citet{2006MNRAS.369..625N} excluding the central 
three softening lengths.  
At high redshift the system is very compact, $r_e = 0.69 \ \rm kpc$, and has a moderate Sersic index of 
$n\approx 2.3$. This is in agreement with the system being flattened and disk-like. At low redshift 
the system is more extended  $r_e = 2.4 \ \rm kpc$, and its Sersic index has increased to $n=3.3$. The galaxy 
is slightly more compact than typical SDSS early-type galaxies at this mass but lies within the observed distribution 
(\citealp{2003MNRAS.343..978S}, see also \citealp{2008ApJ...688..770F}). 
The errors 
given in the figure are bootstrap errors for a fixed projection.  As we have shown before the evolution in 
surface brightness is mainly driven by an evolution in surface density and not by stellar evolution. 

At z=3 the system has a total stellar mass of $M=5.5 \times 10^{10} M_{\odot}$ with an effective radius 
$r_{\mathrm{eff}} = 0.69 \ \rm kpc$ and a corresponding  effective density of 
$\rho_{\mathrm{eff}}= 0.5M/(4/3\pi r_{\mathrm{eff}}^3) = 1.6 \times 10^{10} M_{\odot}$. The projected stellar 
line-of-sight velocity dispersion is $\sigma_{\mathrm{eff}}\approx 240 \ \rm km s^{-1}$. The corresponding 
values at z=0 are  $M=15 \times 10^{10} M_{\odot}$ , $r_{\mathrm{eff}} = 2.4 \
\rm kpc$, $\rho_{\mathrm{eff}}= 1.3 \times 10^{9} M_{\odot}$ 
and $\sigma_{\mathrm{eff}}\approx 190 \ \rm km s^{-1}$, which is a typical dispersion for early-type galaxies at this mass
\citep{1992ApJ...399..462B}. 
From z=3 to z=0 the system accretes about $5.5 \times 10^{10} M_{\odot}$ (see Fig \ref{new}) and 
we can assume for the above scaling relations $\eta =1$ and $\epsilon << 1$ (which is a reasonable 
assumption as most mass is accreted in very small systems and we found the most massive merger since z=3 with a mass ratio 
of 8:1). The z=0 values of the simulated galaxy are close to the simple prediction, however, the size increase as 
well as the decrease in density and dispersion are more moderate. This is however
expected as the real evolution of the system is more complex and there is non-negligible in-situ star formation 
between z=3 and z=1. Still, the simple scaling relations for stellar accretion represent the evolution of the system 
from z=3 to z=0 very well. However, observations of more massive ellipticals than the one presented here indicate 
an even  stronger size increase \citep{2008ApJ...688..770F}. This effect can be expected if the assembly of more massive galaxies
is even more dominated by minor mergers and stellar accretion.

In particular, the drop in velocity 
dispersion is in qualitative agreement with first direct observations by 
\citet{2009ApJ...696L..43C}. Recently, \citet{2009ApJ...697.1290B} have reported a relatively weak evolution 
of the density within fixed 1kpc of only a factor 2-3. This also is in qualitative agreement with our simulation
which shows a decrease of only a factor 1.5 from z=2 to z=0.   
We also note that the stellar population of the system is already evolved at high redshift. At $z=2$ the galaxy 
has a stellar mass of $7.5 \times 10^{10} M_{\odot}$ and a local star
formation rate of only $\approx 6 M_{\odot} \rm yr ^{-1}$
and an average stellar age of $1.6 \ \rm Gyrs$.

\section{Conclusion \& Discussion} 
In this paper we show that the observed size and density evolution of massive spheroids 
agrees with what is to be expected from a high resolution cosmological 
simulation of a system which grows at late times predominantly by minor mergers and accretion of stars. 
We can successfully apply simple scaling relations derived from the virial theorem to demonstrate that  
that the size increase and decrease in density and velocity dispersion is a natural consequence of 
mass assembly by much less massive stellar systems and accretion \citep{2009ApJ...696L..43C} and 
cannot be explained by mass assembly histories dominated by major stellar mergers.  

In the simulation, a first phase $(6 > z > 2)$ dominated by in-situ star formation from inflowing cold gas 
(see e.g. \citealp{2009arXiv0901.2458D} and references therein) produces a massive
and dense stellar system with sizes $r_{1/2} \le 1 \ \rm kpc$. This phase of the formation of 
the cores of ellipticals is followed by an extended phase $(3 > z > 0)$ with little in-situ star formation 
but significant accretion of stellar material. This material can be stripped at larger 
radii and increases the size of the system with time. At the same time the central 
concentration is reduced by dynamical friction from the surviving cores 
(see \citealp{2001ApJ...560..636E}).
The apparent size increase is caused by the initial dominance of the in-situ 
component being heated and, at larger radii, overshadowed ultimately by the accreted stars. 
From z=3 to z=0 the effective radius of the system increases by a factor of 3.5 with a decrease in the 
effective density of more than an order of magnitude and a decrease in velocity dispersion of 20\%, in good agreement 
with predictions from simple scaling relations for the accretion of minor mergers. 


Detailed investigations of dark matter simulations (see e.g. \citealp{2008ApJ...688..789G,2008MNRAS.386..577F,
2008arXiv0812.3154G}) as well 
as recent observations (see e.g. \citealp{2009arXiv0902.1188B}) on the mass assembly 
mechanisms of early-type galaxies and their dark matter halos demonstrate the significance 
of minor mergers. In addition, due to the shape of the mass function, massive systems at high redshift 
are more likely to experience minor mergers, than lower mass galaxies. If the size evolution 
is in general driven by minor mergers we would expect a differential size increase, e.g. 
more massive high redshift systems grow larger than lower mass systems. 
At the same time minor mergers do also play an important role for the gravitational heating of halo gas, 
thereby suppressing the formation of new stars \citep{2009ApJ...697L..38J}. 


A picture of a two phase formation process for massive spheroidal galaxies 
has a number of virtues. In the first dissipative phase at high redshift stars form quickly and 
build the compact progenitor of present-day ellipticals. In fact, it seems of minor importance if the gas is 
funneled to the center through streams or mergers of extended gas dominated disks 
(see e.g. \citealp{2006ApJ...641...90R,2009ApJ...690..802J,2009arXiv0901.2458D}) as long as it happens on 
a short timescale. The stars formed at this early phase are expected to be significantly enriched in 
alpha-elements as expected from observations \citep{2005ApJ...621..673T} and form the 
compact core of the elliptical galaxy. Later on metal-poor stars from smaller systems are accreted 
and form the halo of the galaxy resulting in the observed metalicity gradient. This inside out formation scenario 
is also made plausible by recent observations of \citet{2009ApJ...697.1290B}. There is an important test of 
the picture presented in this paper. If correct, then the outer parts 
of massive giant ellipticals will tend to be old, blue, metal-poor and relatively 
uniform from galaxy to galaxy since they are all composed essentially of the debris 
from tidally destroyed accreted small systems. 

\begin{acknowledgements}
The simulation was performed at the Princeton PICSciE HPC center. This research was supported by the DFG 
cluster of excellence 'Origin and Structure of the Universe'. We thank Marijn Franx, Pieter van Dokkum 
and Ignacio Trujillo for helpful comments on the manuscript.

\end{acknowledgements}


\end{document}